\documentclass[journal,a4paper,twoside,10pt]{IEEEtran}

\makeatletter
\let\l@ENGLISH\l@english
\makeatother

\usepackage[utf8]{inputenc}
\usepackage[english]{babel}
\usepackage{csquotes}

\usepackage[style=ieee,backend=bibtex,url=false,doi=false,isbn=false,sorting=none,firstinits=true,sortcites]{biblatex}
\AtEveryBibitem{%
  \clearfield{language}%
  \clearlist{language}%
}

\defbibheading{bibliography}[\refname]{%
  \section*{#1}%
}

\usepackage{amssymb}
\usepackage{mathtools}
\usepackage{commath}
\usepackage{bm}
\usepackage[mode=math]{siunitx}
\usepackage{graphicx}
\usepackage{tikz}
\usepackage{pgfplots}
\usetikzlibrary{plotmarks}
\pgfplotsset{plot coordinates/math parser=false}

\pgfplotsset{
  every axis legend/.style={
    font = \footnotesize,
  },
  xlabel near ticks,
  ylabel near ticks,
  every axis plot post/.append style = semithick,
  axis line style = thin,
  tick style = thin,
}

\newlength{\figureheight}
\newlength{\figurewidth}
\usepackage{ifpdf}
\ifpdf
\usepackage[pdftex]{hyperref} %
\else
\usepackage[hypertex=true]{hyperref} %
\fi
\usepackage[capitalise]{cleveref} %
\crefname{section}{Sec.}{Secs.}
\crefname{chapter}{Ch.}{Chs.}
\crefname{figure}{Fig.}{Fig.}
\crefname{equation}{\unskip}{\unskip}
\Crefname{equation}{Equation}{Equations}
\addto\captionsenglish{}
\addto\captionsenglish{}
\usepackage{microtype}

\usepackage{array}
\usepackage{booktabs}
\usepackage{pgfplotstable}

\usepackage{flushend}

\newcommand{\tu}[1]{\ensuremath{_{\textrm{#1}}}}

\newcommand{\unitspace}{\ensuremath{\enspace}}

\makeatletter
\long\def\@makecaption#1#2{\ifx\@captype\@IEEEtablestring%
\footnotesize\begin{center}{\normalfont\footnotesize #1}\\
{\normalfont\footnotesize\scshape #2}\end{center}%
\@IEEEtablecaptionsepspace
\else
\@IEEEfigurecaptionsepspace
\setbox\@tempboxa\hbox{\normalfont\footnotesize {#1.}~~ #2}%
\ifdim \wd\@tempboxa >\hsize%
\setbox\@tempboxa\hbox{\normalfont\footnotesize {#1.}~~ }%
\parbox[t]{\hsize}{\normalfont\footnotesize \noindent\unhbox\@tempboxa#2}%
\else
\hbox to\hsize{\normalfont\footnotesize\hfil\box\@tempboxa\hfil}\fi\fi}
\makeatother

\makeatletter
\def\ps@headings{%
\def\@oddhead{\mbox{}\scriptsize\rightmark \hfil }%
\def\@evenhead{\scriptsize \hfil \leftmark\mbox{}}%
\def\@oddfoot{}%
\def\@evenfoot{}}
\makeatother

\makeatletter
\def\ps@IEEEtitlepagestyle{%
\def\@oddhead{\mbox{}\scriptsize\rightmark \hfil }%
\def\@evenhead{\scriptsize \hfil \leftmark\mbox{}}%
\def\@oddfoot{}%
\def\@evenfoot{}}
\makeatother

\begin{document}

\title{Quasiparticle effective temperature in superconducting thin films illuminated at THz frequencies}

\author{\IEEEauthorblockN{T. Guruswamy\IEEEauthorrefmark{1}\IEEEauthorrefmark{2}, D. J. Goldie\IEEEauthorrefmark{1} and S. Withington\IEEEauthorrefmark{1}}\\
\IEEEauthorblockA{{\itshape \IEEEauthorrefmark{1}Quantum Sensors Group\\
Cavendish Laboratory, University of Cambridge\\
JJ Thomson Avenue, Cambridge CB3~0HE, UK} \\
\IEEEauthorrefmark{2}Contact: tg307@mrao.cam.ac.uk}\\
}

\markboth{26th International Symposium on Space Terahertz Technology, Cambridge, MA, 16-18 March, 2015}%
{26th International Symposium on Space Terahertz Technology, Cambridge, MA, 16-18 March, 2015}

\maketitle

\begin{abstract}
The response of superconducting pair-breaking detectors is dependent on the details of the quasiparticle distribution.
In Kinetic Inductance Detectors (KIDs), where both pair breaking and non-pair breaking photons are absorbed simultaneously, calculating the detector response therefore requires knowledge of the often nonequilibrium distributions.
The quasiparticle effective temperature provides a good approximation to these nonequilibrium distributions.
We compare an analytical expression relating absorbed power and the quasiparticle effective temperature in superconducting thin films to full solutions for the nonequilibrium distributions, and find good agreement for a range of materials, absorbed powers, photon frequencies and temperatures typical of KIDs.
This analytical expression allows inclusion of nonequilibrium effects in device models without solving for the detailed distributions.
We also show our calculations of the frequency dependence of the detector response are in agreement with recent experimental measurements of the response of Ta KIDs at THz frequencies.
\end{abstract}

\section{Introduction} \label{sec:intro}
Kinetic Inductance Detectors (KIDs)~\cite{Day2003,Vardulakis2008,Baselmans2012} rely on photons with energy $h\nu \ge 2\Delta$, where $\nu$ is the photon frequency and $\Delta$ the superconducting energy gap, to break Cooper pairs and thereby produce an excess population of quasiparticles.
Sub-gap readout photons are also absorbed, and they too generate an excess population of quasiparticles, which influences the operating characteristics  of the device~\cite{Zmuidzinas2012,Goldie2013}.
Understanding the way in which these distributions are created, and interact, is central to understanding the operation and performance limitations of KIDs.

Once a signal photon is absorbed, the initial high-energy quasiparticles relax towards the superconducting energy gap $\Delta$.
For quasiparticle energies $\Delta < E \ll \Omega_D$ in typical materials, this happens primarily by emitting phonons~\cite{Guruswamy2014}.
In a thin film, these excess phonons can either remain within the superconductor and break pairs themselves if they have sufficient phonon energy $\Omega \ge 2\Delta$, or escape into the substrate.
The energy carried by the escaping phonons is completely lost from the quasiparticle system.
At low temperatures $T \sim 0.1\,T_c$ (where $T_c$ is the superconductor critical temperature), the downconversion process happens very fast ($< \SI{10}{ns}$, as the quasiparticle-phonon scattering lifetime $\tau_s$ is much shorter than the recombination lifetime $\tau_r$~\cite{Kaplan1976,Kozorezov2000}) so it is the long-lived low energy quasiparticles of energy $E \approx \Delta$ which primarily determine the detector response.
We define an associated quasiparticle generation efficiency $\eta$, given by the fraction of the absorbed photon energy that remains detectable as excess low energy quasiparticles.
For very high energy photons ($h\nu \gg \Omega_D$), $\eta = 0.6$ is commonly used~\cite{Zmuidzinas2012,Kurakado1982}, but over the moderate energy range ($2\Delta < h\nu \le 10\Delta$ -- THz spectrum) of signal photons studied here, $\eta$ varies significantly~\cite{Guruswamy2014}.

In \cite{Goldie2013} we introduced a method for calculating the steady state, nonequilibrium quasiparticle and phonon distributions in superconducting thin films with simultaneous above-gap (signal) and sub-gap (readout or probe) photon illumination, by solving the nonlinear Chang \& Scalapino kinetic equations~\cite{Chang1977}.
Using this method, the effect of uniform, constant absorption of sub-gap photons~\cite{Goldie2013,Goldie2014} and moderate energy above-gap photons~\cite{Guruswamy2014}, has been quantified.
A key result of that work was an analytical relationship between the effective quasiparticle temperature $T_N^*$ and absorbed power $P$ (at a single photon frequency $\nu$),
\begin{equation} \label{eq:analytical}
\begin{split}
  P &= \frac{\Sigma_s}{\eta(\nu,P,T_b) \, (1+\tau_l/\tau_{pb})} \times \\
  &\left[
  T_N^* \exp \left( \frac{-2 \Delta(T_N^*)}{k_B T_N^*} \right) - 
  T_b \exp \left( \frac{-2 \Delta(T_b)}{k_B T_b} \right)
  \right] .
\end{split}
\end{equation}
Here the effective quasiparticle temperature $T_N^*$ is defined as the temperature of the thermal distribution which has the same total number of quasiparticles as the steady-state nonequilibrium distribution of interest.
Using $T_N^*$ in equations that assume thermal quasiparticle distributions is often sufficient for calculating key characteristics such as surface impedance~\cite{Goldie2013}.
\Cref{eq:analytical} therefore can be used to calculate the effective temperature from absorbed power (or vice versa) using only a material dependent constant $\Sigma_s$, derived from fitting the effective temperatures of the calculated nonequilibrium distributions; and the material independent but power, temperature and frequency dependent $\eta$, also originally calculated from the nonequilibrium distributions.
$\tau_l$ is the phonon escape time into the substrate; $\tau_{pb}$ is the phonon pair breaking time, which at low temperatures $T \ll T_c$ is equal to the characteristic phonon lifetime $\tau_0^\phi$~\cite{Kaplan1976}; and $T_b$ is the substrate or heat bath temperature.
Our most recent work~\cite{Guruswamy2015pre} has calculated $\Sigma_s$ for a range of common materials (Al, Mo, Ta, Nb, NbN), and also calculated $\eta$ in the sub-gap and above-gap frequency regimes, at a range of temperatures.

In this work, we compare \cref{eq:analytical} with the complete solutions to the Chang \& Scalapino equations at the typical absorbed powers, signal and readout frequencies, temperatures, and for a range of commonly used low-$T_c$ superconductors.
We also compare the frequency dependence of our calculated quasiparticle generation efficiency $\eta$ to recent measurements of Ta KID response~\cite{Neto2014} at THz frequencies.

\section{Results}

\begin{figure}[htb]
  \centering
\makeatletter{}%
\begin{tikzpicture}

\begin{axis}[%
width=0.95092\figurewidth,
height=\figureheight,
at={(0\figurewidth,0\figureheight)},
scale only axis,
separate axis lines,
every outer x axis line/.append style={black},
every x tick label/.append style={font=\color{black}},
xmode=log,
xmin=2e-08,
xmax=20000000,
xminorticks=true,
xlabel={$P\tu{probe}\unitspace (\si{W.m^{-3}})$},
every outer y axis line/.append style={black},
every y tick label/.append style={font=\color{black}},
ymin=0.1,
ymax=0.18,
ylabel={$T_N^* / T_c$},
legend style={at={(0.03,0.97)},anchor=north west,legend cell align=left,align=left,fill=none,draw=none}
]
\addplot [color=blue,only marks,mark=+,mark options={solid}]
  table[row sep=crcr]{%
2e-08	0.100039672851563\\
6.32455532033676e-08	0.100123291015625\\
2e-07	0.100372314453125\\
6.32455532033676e-07	0.101043701171875\\
2e-06	0.1025146484375\\
6.32455532033676e-06	0.104947204589844\\
2e-05	0.108124389648437\\
6.32455532033676e-05	0.11178466796875\\
0.0002	0.115805206298828\\
0.000632455532033676	0.120147171020508\\
0.002	0.124809532165527\\
0.00632455532033676	0.129794883728027\\
0.02	0.13507942199707\\
0.0632455532033676	0.140663080215454\\
0.2	0.146595668792725\\
0.632455532033676	0.153071329593658\\
2	0.160144810676575\\
6.32455532033676	0.167826758623123\\
20	0.176105219125748\\
63.2455532033676	0.185078515410423\\
200	0.19479614675045\\
632.455532033676	0.205338815748692\\
2000	0.216829069107771\\
6324.55532033676	0.229400306493044\\
20000	0.243203012347222\\
63245.5532033676	0.258407568782568\\
200000	0.275272615961731\\
};
\addlegendentry{Al};

\addplot [color=black!50!green,only marks,mark=o,mark options={solid}]
  table[row sep=crcr]{%
0.02	0.102469825744629\\
0.0632455532033676	0.104879837036133\\
0.2	0.108040370941162\\
0.632455532033676	0.111688575744629\\
2	0.115685682296753\\
6.32455532033676	0.120000627040863\\
20	0.124612374305725\\
63.2455532033676	0.129512243270874\\
200	0.134719743728638\\
632.455532033676	0.140272135734558\\
2000	0.146238864660263\\
6324.55532033676	0.152682609111071\\
20000	0.159625560343266\\
63245.5532033676	0.167153247594833\\
200000	0.175281497314572\\
};
\addlegendentry{Ta};

\addplot [color=red,only marks,mark=x,mark options={solid}]
  table[row sep=crcr]{%
0.0632455532033676	0.100188827514648\\
0.2	0.100559844970703\\
0.632455532033676	0.101496696472168\\
2	0.103343152999878\\
6.32455532033676	0.106094245910645\\
20	0.10948034286499\\
63.2455532033676	0.113288702964783\\
200	0.117436975240708\\
632.455532033676	0.121898310184479\\
2000	0.126677769422531\\
6324.55532033676	0.131784662604332\\
20000	0.137224458158016\\
63245.5532033676	0.143006552606821\\
200000	0.149180057495833\\
632455.532033676	0.15579976990819\\
2000000	0.16292365718633\\
6324555.32033676	0.170608679018915\\
20000000	0.178922357708216\\
};
\addlegendentry{Nb};

\addplot [color=black,only marks,mark=diamond,mark options={solid}]
  table[row sep=crcr]{%
2e-08	0.100103149414063\\
6.32455532033676e-08	0.100313720703125\\
2e-07	0.10089599609375\\
6.32455532033676e-07	0.102222595214844\\
2e-06	0.104513854980469\\
6.32455532033676e-06	0.107596435546875\\
2e-05	0.111195602416992\\
6.32455532033676e-05	0.115169525146484\\
0.0002	0.119472160339355\\
0.000632455532033676	0.124104347229004\\
0.002	0.129080085754395\\
0.00632455532033676	0.134410266876221\\
0.02	0.140073671340942\\
0.0632455532033676	0.146078505516052\\
0.2	0.152451672554016\\
0.632455532033676	0.159373905658722\\
2	0.166954917907715\\
6.32455532033676	0.175186592340469\\
20	0.184089664816856\\
63.2455532033676	0.193719844818115\\
200	0.204173053205013\\
632.455532033676	0.215563116967678\\
2000	0.22802346214652\\
6324.55532033676	0.241705069243908\\
20000	0.256779454648495\\
63245.5532033676	0.27351835437119\\
200000	0.293523845747113\\
};
\addlegendentry{Mo};

\addplot [color=red!50!green,only marks,mark=asterisk,mark options={solid}]
  table[row sep=crcr]{%
0.2	0.100105476379395\\
0.632455532033676	0.10032096862793\\
2	0.100914344787598\\
6.32455532033676	0.102260036468506\\
20	0.10456974029541\\
63.2455532033676	0.107664642333984\\
200	0.111276688575745\\
632.455532033676	0.115257129669189\\
2000	0.119564318656921\\
6324.55532033676	0.124198689460754\\
20000	0.129171714782715\\
63245.5532033676	0.134488830566406\\
200000	0.140149556398392\\
632455.532033676	0.146147883236408\\
2000000	0.152534007430076\\
6324555.32033676	0.159368042051792\\
20000000	0.166720484569669\\
};
\addlegendentry{NbN};

\addplot [color=blue,solid,forget plot]
  table[row sep=crcr]{%
2.16306349842986e-08	0.100039672851563\\
6.82570065550972e-08	0.100123291015625\\
2.15205970687899e-07	0.100372314453125\\
6.79306074074763e-07	0.101043701171875\\
2.14470998937518e-06	0.1025146484375\\
6.7760666837686e-06	0.104947204589844\\
2.13996922039768e-05	0.108124389648437\\
6.75486534748118e-05	0.11178466796875\\
0.000213253237666189	0.115805206298828\\
0.00067298821041837	0.120147171020508\\
0.00212266226919512	0.124809532165527\\
0.00668864724780977	0.129794883728027\\
0.0210281968253965	0.13507942199707\\
0.0660600401104844	0.140663080215454\\
0.207194023792147	0.146595668792725\\
0.64913277471647	0.153071329593658\\
2.02725002113767	0.160144810676575\\
6.30824950459554	0.167826758623123\\
19.5291844922904	0.176105219125748\\
60.1711607918258	0.185078515410423\\
184.228661439454	0.19479614675045\\
559.845332102426	0.205338815748692\\
1686.85029116535	0.216829069107771\\
5032.42287080879	0.229400306493044\\
14839.9292465096	0.243203012347222\\
43164.2573161781	0.258407568782568\\
123259.79951695	0.275272615961731\\
};
\addplot [color=black!50!green,dashed,forget plot]
  table[row sep=crcr]{%
0.0204960002064443	0.102469825744629\\
0.0647359563188236	0.104879837036133\\
0.204427929264562	0.108040370941162\\
0.645444363905049	0.111688575744629\\
2.03662542934842	0.115685682296753\\
6.42284412120426	0.120000627040863\\
20.2618091892575	0.124612374305725\\
63.8621312279839	0.129512243270874\\
201.127154717443	0.134719743728638\\
631.897116822207	0.140272135734558\\
1980.94512775042	0.146238864660263\\
6195.45449675911	0.152682609111071\\
19301.2398075497	0.159625560343266\\
59976.1338491509	0.167153247594833\\
185552.039877709	0.175281497314572\\
};
\addplot [color=red,dash pattern=on 1pt off 3pt on 3pt off 3pt,forget plot]
  table[row sep=crcr]{%
0.0649021448876768	0.100188827514648\\
0.205153737246774	0.100559844970703\\
0.648360879381181	0.101496696472168\\
2.04785962887362	0.103343152999878\\
6.46025347621315	0.106094245910645\\
20.3978908810887	0.10948034286499\\
64.4171328612804	0.113288702964783\\
203.570648884667	0.117436975240708\\
642.485497275842	0.121898310184479\\
2026.28477764706	0.126677769422531\\
6386.88148292925	0.131784662604332\\
20101.6116739221	0.137224458158016\\
63099.1695737573	0.143006552606821\\
197596.270726771	0.149180057495833\\
616957.751980552	0.15579976990819\\
1919763.29642936	0.16292365718633\\
5950543.59236408	0.170608679018915\\
18367280.06888	0.178922357708216\\
};
\addplot [color=black,dotted,forget plot]
  table[row sep=crcr]{%
2.1823368174881e-08	0.100103149414063\\
6.87998161449497e-08	0.100313720703125\\
2.17665896197762e-07	0.10089599609375\\
6.87584912965239e-07	0.102222595214844\\
2.17178795780011e-06	0.104513854980469\\
6.85826467260208e-06	0.107596435546875\\
2.1649575562308e-05	0.111195602416992\\
6.83516989593433e-05	0.115169525146484\\
0.000215734600502566	0.119472160339355\\
0.000680604773178456	0.124104347229004\\
0.00214556505926684	0.129080085754395\\
0.00675602745772827	0.134410266876221\\
0.0212191797310334	0.140073671340942\\
0.0665768166743269	0.146078505516052\\
0.208466696887386	0.152451672554016\\
0.65147443255548	0.159373905658722\\
2.02964561906005	0.166954917907715\\
6.29543279853958	0.175186592340469\\
19.4226931342131	0.184089664816856\\
59.5358958668201	0.193719844818115\\
181.14474552193	0.204173053205013\\
546.433979724886	0.215563116967678\\
1631.93841109544	0.22802346214652\\
4817.07222240933	0.241705069243908\\
14023.3400307652	0.256779454648495\\
40058.4199202151	0.27351835437119\\
109195.521194668	0.293523845747113\\
};
\addplot [color=red!50!green,solid,forget plot]
  table[row sep=crcr]{%
0.20338582620078	0.100105476379395\\
0.642328116420676	0.10032096862793\\
2.03071310375254	0.100914344787598\\
6.41570692500886	0.102260036468506\\
20.2634993206904	0.10456974029541\\
63.9911866051443	0.107664642333984\\
202.235455024509	0.111276688575745\\
638.596821631724	0.115257129669189\\
2015.81664844573	0.119564318656921\\
6361.81257828864	0.124198689460754\\
20070.0589869184	0.129171714782715\\
63251.4935345553	0.134488830566406\\
199000.318095961	0.140149556398392\\
624027.757961286	0.146147883236408\\
1951286.85181988	0.152534007430076\\
6080969.15207009	0.159368042051792\\
18881559.7262696	0.166720484569669\\
};
\end{axis}
\end{tikzpicture}

  \caption{Quasiparticle effective temperature $T_N^*$ against absorbed sub-gap power $P\tu{probe}$ for Al, Ta, Nb, Mo, and NbN, from full nonequilibrium calculation (markers) and analytical expression (lines). Calculated with $h\nu_p = \SI{16}{\micro eV}$, $T_b = 0.1\,T_c$ and $\tau_l / \tau_0^\phi = 1$.}
  \label{fig:varying_material}
\end{figure}
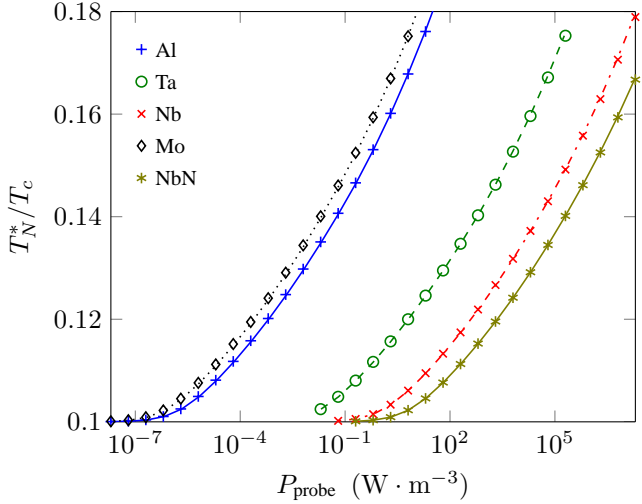

In \cref{fig:varying_material}, we compare the quasiparticle effective temperature $T_N^*$ calculated from \cref{eq:analytical} (lines) -- for sub-gap readout frequency $\nu_p$ and varying readout power $P\tu{probe}$ -- to $T_N^*$ calculated from the full nonequilibrium distributions for Al, Ta, Nb, Mo, and NbN (markers), using the values for $\Sigma_s$ tabulated in \cite{Guruswamy2015pre}.
There is excellent agreement for all readout powers considered.
The material-dependent constant $\Sigma_s$ scales with the zero-temperature superconducting gap energy of the material, so for the same absorbed power, a greater effective temperature change is seen in Mo than NbN.

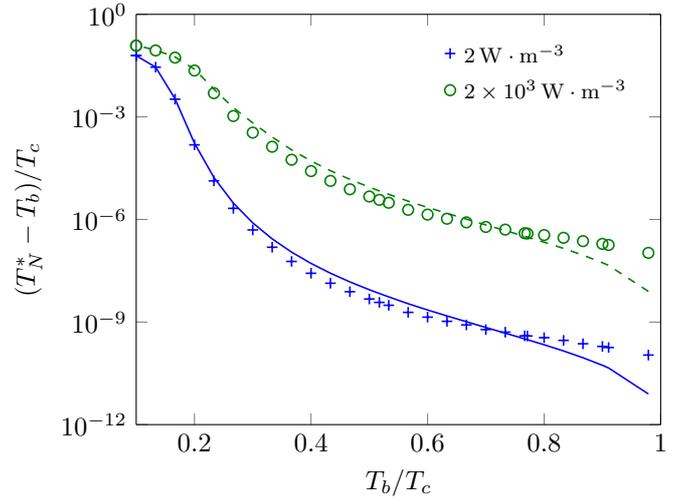
\begin{figure}[htb]
  \centering
\makeatletter{}%
\begin{tikzpicture}

\begin{axis}[%
width=0.95092\figurewidth,
height=\figureheight,
at={(0\figurewidth,0\figureheight)},
scale only axis,
separate axis lines,
every outer x axis line/.append style={black},
every x tick label/.append style={font=\color{black}},
xmin=0.1,
xmax=1,
xlabel={$T_b / T_c$},
every outer y axis line/.append style={black},
every y tick label/.append style={font=\color{black}},
ymode=log,
ymin=1e-12,
ymax=1,
yminorticks=true,
ylabel={$(T_N^* - T_b)/T_c$},
legend style={at={(0.97,0.97)},anchor=north east,legend cell align=left,align=left,fill=none,draw=none}
]
\addplot [color=blue,only marks,mark=+,mark options={solid}]
  table[row sep=crcr]{%
0.978975717387908	1.08269504215164e-10\\
0.910652705694886	1.80223654025346e-10\\
0.9	1.93831528209195e-10\\
0.866666666666666	2.32054673984412e-10\\
0.833333333333334	2.91038317692857e-10\\
0.799999999999998	3.49245924976523e-10\\
0.771287709899638	3.90585311255595e-10\\
0.76666666666731	3.94938878411356e-10\\
0.733333333333345	5.05824517018286e-10\\
0.7	6.03031306501948e-10\\
0.666666666666785	8.24608410532802e-10\\
0.633333333333961	1.05064821623657e-09\\
0.599999999998738	1.38999899705037e-09\\
0.566666666667764	1.91968862649188e-09\\
0.533333333333637	3.09199098290983e-09\\
0.517256735831825	3.75751652051336e-09\\
0.500000000000047	4.72064129170885e-09\\
0.46666666666667	7.71445518176906e-09\\
0.433333333333336	1.36609499373563e-08\\
0.399999999999997	2.67103314174727e-08\\
0.366666666666662	5.92134893460397e-08\\
0.333333333333334	1.53994187713914e-07\\
0.300000000000001	4.94308769764152e-07\\
0.26666666666668	2.11189190544288e-06\\
0.233333333373164	1.34528800867971e-05\\
0.199999999993975	0.000152537822718786\\
0.166666666666375	0.00330020984013299\\
0.133333334819895	0.0284730024333371\\
0.0999999997825829	0.0620552586159968\\
0.1	0.0620552587509156\\
};
\addlegendentry{$\SI{2}{W.m^{-3}}$};

\addplot [color=black!50!green,only marks,mark=o,mark options={solid}]
  table[row sep=crcr]{%
0.978975717387908	1.05534157989951e-07\\
0.910652705694886	1.79370265417517e-07\\
0.9	1.9350147336404e-07\\
0.866666666666666	2.3242280196226e-07\\
0.833333333333334	2.91360871883913e-07\\
0.799999999999998	3.49292531697144e-07\\
0.771287709899638	3.88181105782735e-07\\
0.76666666666731	3.95541428644656e-07\\
0.733333333333345	5.04420216430305e-07\\
0.7	6.03952212341413e-07\\
0.666666666666785	8.22897224432664e-07\\
0.633333333333961	1.04980407452887e-06\\
0.599999999998738	1.38872070248208e-06\\
0.566666666667764	1.91204288658373e-06\\
0.533333333333637	3.07797764736744e-06\\
0.517256735831825	3.74602107392019e-06\\
0.500000000000047	4.69665974381929e-06\\
0.46666666666667	7.64613350235591e-06\\
0.433333333333336	1.3450713595523e-05\\
0.399999999999997	2.59590055793617e-05\\
0.366666666666662	5.58154750614738e-05\\
0.333333333333334	0.000134121968100486\\
0.300000000000001	0.000347113553434624\\
0.26666666666668	0.00105960796276736\\
0.233333333373164	0.00494983314310564\\
0.199999999993975	0.0226810786121211\\
0.166666666666375	0.0541982911526209\\
0.133333334819895	0.087715247655355\\
0.0999999997825829	0.121184129004456\\
0.1	0.121184135377407\\
};
\addlegendentry{$\SI{2E3}{W.m^{-3}}$};

\addplot [color=blue,solid,forget plot]
  table[row sep=crcr]{%
0.978975717387908	7.86649850330949e-12\\
0.910652705694886	4.53579554711098e-11\\
0.9	5.43647409139112e-11\\
0.866666666666666	9.04038837316762e-11\\
0.833333333333334	1.42485706812197e-10\\
0.799999999999998	2.16623297030773e-10\\
0.771287709899638	3.04196527663436e-10\\
0.76666666666731	3.20960208301516e-10\\
0.733333333333345	4.73983302783715e-10\\
0.7	6.91978000658424e-10\\
0.666666666666785	1.02276134684331e-09\\
0.633333333333961	1.50827287370358e-09\\
0.599999999998738	2.25337697600954e-09\\
0.566666666667764	3.43130744823112e-09\\
0.533333333333637	5.44674483080788e-09\\
0.517256735831825	6.83891686110908e-09\\
0.500000000000047	8.82022786142165e-09\\
0.46666666666667	1.49358716490144e-08\\
0.433333333333336	2.68010180758639e-08\\
0.399999999999997	5.18652944323694e-08\\
0.366666666666662	1.10876024737295e-07\\
0.333333333333334	2.70870846834329e-07\\
0.300000000000001	7.94763225597689e-07\\
0.26666666666668	3.02454796326142e-06\\
0.233333333373164	1.69083386146707e-05\\
0.199999999993975	0.000169759764796146\\
0.166666666666375	0.00337639063525669\\
0.133333334819895	0.0284536339219649\\
0.0999999997825829	0.0619056816119606\\
0.1	0.0619056814925769\\
};
\addplot [color=black!50!green,dashed,forget plot]
  table[row sep=crcr]{%
0.978975717387908	7.86660801285935e-09\\
0.910652705694886	4.5357986411308e-08\\
0.9	5.43640756058926e-08\\
0.866666666666666	9.04026588748617e-08\\
0.833333333333334	1.4248181856687e-07\\
0.799999999999998	2.16614292214252e-07\\
0.771287709899638	3.04181877010816e-07\\
0.76666666666731	3.20943974729607e-07\\
0.733333333333345	4.73950292592539e-07\\
0.7	6.91916663403217e-07\\
0.666666666666785	1.0226313586326e-06\\
0.633333333333961	1.50801089788905e-06\\
0.599999999998738	2.25281856965709e-06\\
0.566666666667764	3.43004110304863e-06\\
0.533333333333637	5.44342089550436e-06\\
0.517256735831825	6.83364266770073e-06\\
0.500000000000047	8.81134008304444e-06\\
0.46666666666667	1.49092578989256e-05\\
0.433333333333336	2.67093023085098e-05\\
0.399999999999997	5.14881683029871e-05\\
0.366666666666662	0.000108932943488959\\
0.333333333333334	0.000257492952665667\\
0.300000000000001	0.000670953523394074\\
0.26666666666668	0.0018590496726666\\
0.233333333373164	0.0064820397003561\\
0.199999999993975	0.0248471636311351\\
0.166666666666375	0.056397399368525\\
0.133333334819895	0.0898305574736188\\
0.0999999997825829	0.123227855971619\\
0.1	0.123227859681333\\
};
\end{axis}
\end{tikzpicture}%

  \caption{Quasiparticle effective temperature difference $T_N^* - T_b$ against substrate temperature $T_b$ for selected absorbed sub-gap powers $P\tu{probe}$, from full nonequilibrium calculation (markers) and analytical expression (lines). Calculated for Al, with $h\nu_p = \SI{16}{\micro eV}$ and $\tau_l/\tau_0^\phi = 1$.}
  \label{fig:varying_temp}
\end{figure}

In \cref{fig:varying_temp} the effective temperature calculated from \cref{eq:analytical} (lines) is compared to $T_N^*$ calculated from the nonequilibrium distributions (markers) when varying the substrate temperature $T_b$.
As the substrate temperature increases, the same absorbed power causes a smaller increase in the quasiparticle effective temperature $T_N^*$, as temperature is a nonlinear function of total quasiparticle number and energy.
\Cref{eq:analytical} is in reasonable agreement with the full calculation until $T_b \approx 0.8\,T_c$, where $k_B T \approx \Delta(T)$.

\begin{figure}[htb]
  \centering
\makeatletter{}%
\begin{tikzpicture}

\begin{axis}[%
width=0.95092\figurewidth,
height=\figureheight,
at={(0\figurewidth,0\figureheight)},
scale only axis,
separate axis lines,
every outer x axis line/.append style={black},
every x tick label/.append style={font=\color{black}},
xmode=log,
xmin=2e-08,
xmax=2000,
xminorticks=true,
xlabel={$P\tu{probe}\unitspace (\si{W.m^{-3}})$},
every outer y axis line/.append style={black},
every y tick label/.append style={font=\color{black}},
ymin=0.1,
ymax=0.25,
ylabel={$T_N^* / T_c$},
legend style={at={(0.03,0.97)},anchor=north west,legend cell align=left,align=left,fill=none,draw=none}
]
\addplot [color=blue,only marks,mark=+,mark options={solid}]
  table[row sep=crcr]{%
2e-08	0.100043334960938\\
2e-07	0.100408325195313\\
2e-06	0.10269287109375\\
2e-05	0.10843505859375\\
0.0002	0.116199798583984\\
0.002	0.125324363708496\\
0.02	0.13590163230896\\
0.2	0.148048577308655\\
2	0.162055258750916\\
20	0.178562636375427\\
200	0.197993168234825\\
2000	0.221184135377407\\
2199.999999998	0.221494173592988\\
};
\addlegendentry{$\tau_l/\tau_0^\phi = 1$};

\addplot [color=black!50!green,only marks,mark=o,mark options={solid}]
  table[row sep=crcr]{%
2e-08	0.100125732421875\\
2e-07	0.101064453125\\
2e-06	0.105009765625\\
2e-05	0.111881713867187\\
0.0002	0.120302276611328\\
0.002	0.130129280090332\\
0.02	0.141618623733521\\
0.2	0.155051317214966\\
2	0.170594102144241\\
20	0.188965525627136\\
200	0.21097677052021\\
2000	0.237734171748162\\
2199.999999998	0.238100613507956\\
};
\addlegendentry{$\tau_l/\tau_0^\phi = 5$};

\addplot [color=red,only marks,mark=x,mark options={solid}]
  table[row sep=crcr]{%
2e-08	0.100224609375\\
2e-07	0.10172119140625\\
2e-06	0.106599731445313\\
2e-05	0.113946990966797\\
0.0002	0.122723999023438\\
0.002	0.132968692779541\\
0.02	0.144998922348023\\
0.2	0.159189929962158\\
2	0.17578058719635\\
20	0.19532868295908\\
200	0.218890690207482\\
2000	0.247717829197645\\
};
\addlegendentry{$\tau_l/\tau_0^\phi = 10$};

\addplot [color=blue,solid,forget plot]
  table[row sep=crcr]{%
2.14343603283001e-08	0.100043334960938\\
2.1523192887683e-07	0.100408325195313\\
2.14976285338857e-06	0.10269287109375\\
2.14238607905321e-05	0.10843505859375\\
0.000213638788855529	0.116199798583984\\
0.00212814151293306	0.125324363708496\\
0.0211363928487095	0.13590163230896\\
0.208618523789194	0.148048577308655\\
2.04252376591919	0.162055258750916\\
19.7093475716247	0.178562636375427\\
186.126085562564	0.197993168234825\\
1711.85400812931	0.221184135377407\\
2199.999999998	0.222196762782158\\
};
\addplot [color=black!50!green,dashed,forget plot]
  table[row sep=crcr]{%
2.15462765391531e-08	0.100125732421875\\
2.15007950881514e-07	0.101064453125\\
2.14594555981662e-06	0.105009765625\\
2.14085570447285e-05	0.111881713867187\\
0.000213541219933207	0.120302276611328\\
0.002127672037231	0.130129280090332\\
0.0211574261289962	0.141618623733521\\
0.209346908120343	0.155051317214966\\
2.05404519412093	0.170594102144241\\
19.9133735271218	0.188965525627136\\
189.662026847221	0.21097677052021\\
1768.90508546541	0.237734171748162\\
2199.999999998	0.238731487061098\\
};
\addplot [color=red,dash pattern=on 1pt off 3pt on 3pt off 3pt,forget plot]
  table[row sep=crcr]{%
2.14739201551136e-08	0.100224609375\\
2.14799660945429e-07	0.10172119140625\\
2.14436456152094e-06	0.106599731445313\\
2.14007371560852e-05	0.113946990966797\\
0.000213450295654909	0.122723999023438\\
0.00212674510095999	0.132968692779541\\
0.0211537208651802	0.144998922348023\\
0.209567547525171	0.159189929962158\\
2.05926963465959	0.17578058719635\\
20.0020836615199	0.19532868295908\\
191.16515704128	0.218890690207482\\
1791.49216520674	0.247717829197645\\
};
\end{axis}
\end{tikzpicture}

  \caption{Quasiparticle effective temperature $T_N^*$ against absorbed sub-gap power $P\tu{probe}$ for selected phonon escape time ratios $\tau_l/\tau_0^\phi$, from full nonequilibrium calculation (markers) and analytical expression (lines). Calculated for Al, with $h\nu_p = \SI{16}{\micro eV}$, and $T_b = 0.1\,T_c$.}
  \label{fig:varying_taul}
\end{figure}
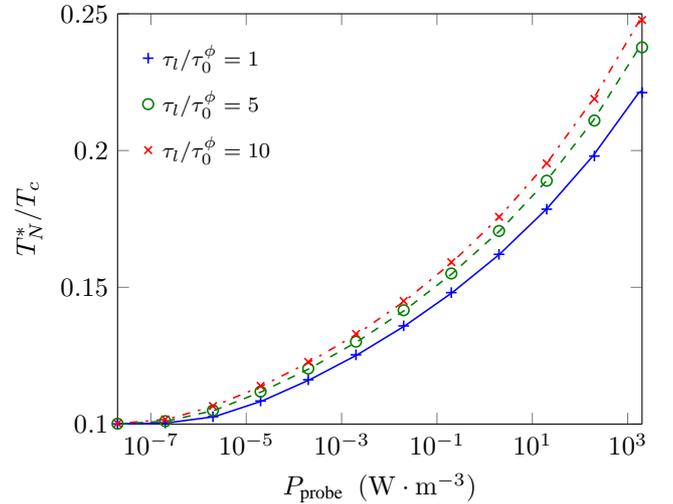

\cref{fig:varying_taul} compares the effective temperatures as a function of absorbed sub-gap power for selected phonon escape time ratios. The effect of phonon trapping is correctly taken into account by \cref{eq:analytical} as shown by the close agreement.
Increasing the phonon escape time $\tau_l$ causes a greater increase in quasiparticle effective temperature for the same absorbed power, as the probability for phonons to escape instead of breaking Cooper pairs is a function of $\tau_l/\tau_{pb}$.

\begin{figure}[htb]
  \centering
\makeatletter{}%
\begin{tikzpicture}

\begin{axis}[%
width=0.95092\figurewidth,
height=\figureheight,
at={(0\figurewidth,0\figureheight)},
scale only axis,
separate axis lines,
every outer x axis line/.append style={black},
every x tick label/.append style={font=\color{black}},
xmode=log,
xmin=0.2,
xmax=2000,
xminorticks=true,
xlabel={$P\unitspace (\si{W.m^{-3}})$},
every outer y axis line/.append style={black},
every y tick label/.append style={font=\color{black}},
ymin=0.14,
ymax=0.23,
ylabel={$T_N^* / T_c$},
legend style={at={(0.03,0.97)},anchor=north west,legend cell align=left,align=left,fill=none,draw=none}
]
\addplot [color=blue,only marks,mark=+,mark options={solid}]
  table[row sep=crcr]{%
0.2	0.145980916023254\\
0.632455532033676	0.152955598831177\\
2	0.160612187385559\\
6.32455532033676	0.169053238630295\\
20	0.178401412367821\\
20	0.178401561379433\\
63.2455532033676	0.188806284070015\\
200	0.200449595451355\\
632.455532033676	0.213557248264551\\
2000	0.22841150701046\\
2000	0.228411860615015\\
};
\addlegendentry{above-gap ($P\tu{signal}$)};

\addplot [color=black!50!green,only marks,mark=o,mark options={solid}]
  table[row sep=crcr]{%
0.0632455532033676	0.140663080215454\\
0.2	0.146595668792725\\
0.632455532033676	0.153071329593658\\
2	0.160144810676575\\
6.32455532033676	0.167826758623123\\
20	0.176105219125748\\
63.2455532033676	0.185078515410423\\
200	0.19479614675045\\
632.455532033676	0.205338815748692\\
2000	0.216829069107771\\
2199.98	0.217410399641453\\
};
\addlegendentry{sub-gap ($P\tu{probe}$)};

\addplot [color=blue,solid,forget plot]
  table[row sep=crcr]{%
0.210778523998843	0.145980916023254\\
0.664265385253255	0.152955598831177\\
2.09171784634967	0.160612187385559\\
6.57911016039555	0.169053238630295\\
20.6614083601046	0.178401412367821\\
20.6614090170486	0.178401561379433\\
64.7586621890685	0.188806284070015\\
202.495382770189	0.200449595451355\\
631.515427030509	0.213557248264551\\
1963.93756825825	0.22841150701046\\
1963.93845750462	0.228411860615015\\
};
\addplot [color=black!50!green,dashed,forget plot]
  table[row sep=crcr]{%
0.0661070232348564	0.140663080215454\\
0.20732517343315	0.146595668792725\\
0.649173667567638	0.153071329593658\\
2.02770149295213	0.160144810676575\\
6.31195264496005	0.167826758623123\\
19.5620095673767	0.176105219125748\\
60.3150110006664	0.185078515410423\\
184.850008095293	0.19479614675045\\
562.599888649214	0.205338815748692\\
1698.78836957245	0.216829069107771\\
2199.98	0.218690870133636\\
};
\end{axis}
\end{tikzpicture}

  \caption{Quasiparticle effective temperature $T_N^*$ against absorbed power $P$ in the sub-gap (dashed line, $\circ$ markers) and above-gap (solid line, $+$ markers) cases, from full nonequilibrium calculation (markers) and analytical expression (lines). Calculated for Al, with $h\nu_p = \SI{16}{\micro eV}$, $h\nu_s = 10\Delta$, and $T_b = 0.1\,T_c$.}
  \label{fig:varying_freq}
\end{figure}
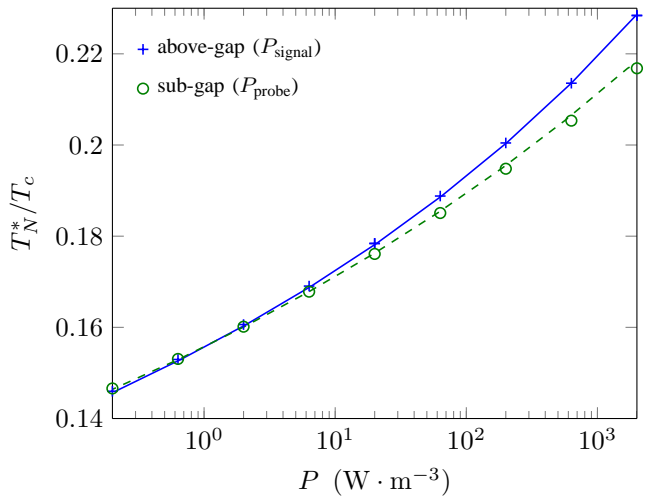

\Cref{fig:varying_freq} compares the effective temperatures as a function of absorbed power when the photons are sub-gap (dashed line, $\circ$ markers -- frequency $h\nu_p$, absorbed power $P\tu{probe}$) and above-gap frequency (solid line, $+$ markers -- frequency $\nu_s$, absorbed power $P\tu{signal}$), showing the analytical expression \cref{eq:analytical} (lines) reproduces the quasiparticle effective temperature from the full calculation (markers).
At the same absorbed power, all parameters of \cref{eq:analytical} are identical between the sub-gap and above-gap cases except for $\eta$.
For direct pair breaking, the required $\eta$ is calculated from the full nonequilibrium distributions using a set of modified Rothwarf-Taylor rate equations~\cite{Guruswamy2014,Guruswamy2015pre}.
In the sub-gap case $\eta = \eta_{2\Delta}$, the fraction of phonons escaping the thin film which have energy $\Omega \ge 2\Delta$.
$\eta$ is constant with absorbed above-gap power~\cite{Guruswamy2014}, while it decreases with absorbed sub-gap power~\cite{Goldie2013,Guruswamy2015pre}.
For the same absorbed power, above-gap power is more efficiently converted into excess quasiparticles than sub-gap power -- so the above-gap absorbed power results in a higher effective temperature than the sub-gap absorbed power.

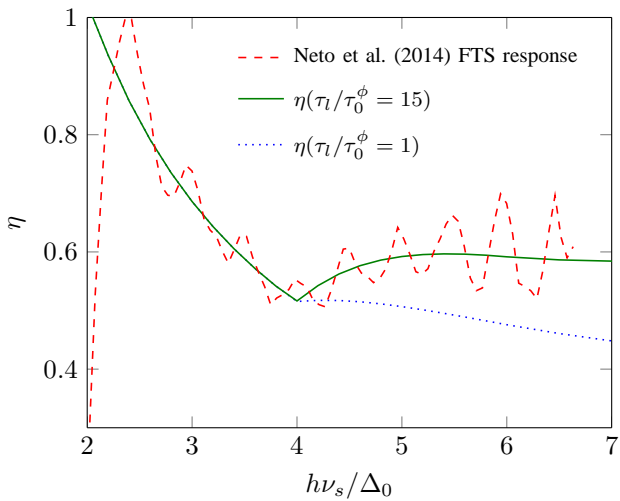
\begin{figure}[htb]
  \centering
\makeatletter{}%
\begin{tikzpicture}

\begin{axis}[%
width=0.95092\figurewidth,
height=\figureheight,
at={(0\figurewidth,0\figureheight)},
scale only axis,
separate axis lines,
every outer x axis line/.append style={black},
every x tick label/.append style={font=\color{black}},
xmin=2,
xmax=7,
xlabel={$h\nu_s / \Delta_0$},
every outer y axis line/.append style={black},
every y tick label/.append style={font=\color{black}},
ymin=0.3,
ymax=1,
ylabel={$\eta$},
legend style={at={(0.97,0.97)},anchor=north east,legend cell align=left,align=left,fill=none,draw=none},
legend reversed=true,
]
\addplot [color=blue,dotted]
  table[row sep=crcr]{%
2	1.02377315507265\\
2.2	0.935430279949\\
2.4	0.857475785612447\\
2.6	0.791521086218347\\
2.8	0.734986896255214\\
3	0.685985383682893\\
3.2	0.643112457518734\\
3.4	0.605279611623148\\
3.6	0.571653405579951\\
3.8	0.541567350768594\\
4	0.515474742646116\\
4.2	0.517328789437557\\
4.4	0.517104838912488\\
4.6	0.514955970650998\\
4.8	0.511345815257119\\
5	0.506667460264257\\
5.2	0.501229491982527\\
5.4	0.495267218107639\\
5.6	0.48895778898126\\
5.8	0.482433778510316\\
6	0.475794765338131\\
6.5	0.460740767149483\\
7	0.448163603375622\\
7.5	0.437294372218628\\
};
\addlegendentry{$\eta(\tau_l/\tau_0^\phi = 1)$};

\addplot [color=black!50!green,solid]
  table[row sep=crcr]{%
2	1.02377003654959\\
2.2	0.935408457933258\\
2.4	0.857456157403948\\
2.6	0.791497244277643\\
2.8	0.734961084646642\\
3	0.68596308130382\\
3.2	0.643089820030348\\
3.4	0.605260470592094\\
3.6	0.571634492696028\\
3.8	0.541549052157363\\
4	0.516367200639958\\
4.2	0.542489221969793\\
4.4	0.562329721717823\\
4.6	0.576475690565134\\
4.8	0.586090436649399\\
5	0.592161451864502\\
5.2	0.595478464716172\\
5.4	0.596658164219955\\
5.6	0.596179165140887\\
5.8	0.594412857300869\\
6	0.591650386149214\\
6.5	0.586428742357576\\
7	0.584306209165358\\
7.5	0.583277955330868\\
};
\addlegendentry{$\eta(\tau_l/\tau_0^\phi = 15)$};

\addplot [color=red,dashed]
  table[row sep=crcr]{%
2.01871141975309	0.28189387848\\
2.07638888888889	0.52774553041\\
2.13406635802469	0.70017248054\\
2.19174382716049	0.85919395154\\
2.2494212962963	0.9077510524\\
2.3070987654321	0.95772733433\\
2.3647762345679	1\\
2.4224537037037	0.99699321332\\
2.48013117283951	0.93637372446\\
2.53780864197531	0.88528736565\\
2.59548611111111	0.8446466906\\
2.65316358024691	0.7652736316\\
2.71084104938272	0.7115244763\\
2.76851851851852	0.69667267416\\
2.82619598765432	0.69502237788\\
2.88387345679012	0.71891622833\\
2.94155092592593	0.74794853105\\
2.99922839506173	0.73803901252\\
3.05690586419753	0.70371286908\\
3.11458333333333	0.65537300746\\
3.17226080246914	0.6371341055\\
3.22993827160494	0.6270893103\\
3.28761574074074	0.60152616781\\
3.34529320987654	0.58068811388\\
3.40297067901235	0.60438042958\\
3.46064814814815	0.62940825954\\
3.51832561728395	0.62654981521\\
3.57600308641975	0.59133801655\\
3.63368055555556	0.56489240829\\
3.69135802469136	0.540977416\\
3.74903549382716	0.50982843802\\
3.80671296296296	0.52139749337\\
3.86439043209877	0.52713423844\\
3.92206790123457	0.53582910056\\
3.97974537037037	0.55344190771\\
4.03742283950617	0.5460501545\\
4.09510030864198	0.54014454639\\
4.15277777777778	0.51795372051\\
4.21045524691358	0.51002823916\\
4.26813271604938	0.5064359279\\
4.32581018518519	0.5343035879\\
4.38348765432099	0.5688192744\\
4.44116512345679	0.60478949236\\
4.49884259259259	0.60642906582\\
4.55652006172839	0.58011650771\\
4.6141975308642	0.56386242011\\
4.671875	0.55788762181\\
4.7295524691358	0.54758515406\\
4.78722993827161	0.55423042434\\
4.84490740740741	0.57089023885\\
4.90258487654321	0.60171035249\\
4.96026234567901	0.64177129754\\
5.01793981481481	0.62234720554\\
5.07561728395062	0.59608795088\\
5.13329475308642	0.56619397665\\
5.19097222222222	0.56195464982\\
5.24864969135802	0.57136979823\\
5.30632716049383	0.60132379833\\
5.36400462962963	0.62227523929\\
5.42168209876543	0.64753058836\\
5.47935956790123	0.66275528525\\
5.53703703703704	0.65223509952\\
5.59471450617284	0.60948505447\\
5.65239197530864	0.55550698081\\
5.71006944444444	0.53390923441\\
5.76774691358025	0.53854765713\\
5.82542438271605	0.59686832439\\
5.88310185185185	0.65307731368\\
5.94077932098765	0.69906627483\\
5.99845679012346	0.68272778681\\
6.05613425925926	0.630496473\\
6.11381172839506	0.54658946039\\
6.17148919753086	0.54124561157\\
6.22916666666667	0.53373614524\\
6.28684413580247	0.52010365038\\
6.34452160493827	0.56852008828\\
6.40219907407407	0.63559879988\\
6.45987654320988	0.6961559574\\
6.51755401234568	0.62480677659\\
6.57523148148148	0.58996883532\\
6.63290895061728	0.60897225283\\
};
\addlegendentry{Neto et al. (2014) FTS response};

\end{axis}
\end{tikzpicture}%

  \caption{Quasiparticle generation efficiency $\eta$ as a function of signal frequency $\nu_s$ as calculated from full nonequilibrium distributions at two different phonon escape time ratios (dotted and solid lines), compared to measured FTS spectral response of a Ta KID from \cite{Neto2014} (dashed line).}
  \label{fig:eta}
\end{figure}

\Citeauthor{Neto2014} have recently measured the frequency dependence of the response of a Ta KID, on a SiN membrane to increase the thermal isolation (and hence $\tau_l$), using a Fourier Transform spectrometer (FTS)~\cite{Neto2014}.
Their data analysis of the normalized response took account of antenna effects and band-defining filters, but does not remove the superconducting absorption efficiency, which cuts on after $h\nu_s = 2\Delta$.
The remaining oscillations of period $\sim 0.5\Delta$ are interpreted as due to standing waves and not intrinsic to the KID~\cite{DeVisserEmail}.
\Cref{fig:eta} shows their response measurements (dashed line) along with our calculations of the quasiparticle generation efficiency $\eta$ at two different trapping factors (dotted line and solid line).
Our high trapping factor calculation of $\eta$ ($\tau_l/\tau_0^\phi = 15$, solid line) shows excellent agreement with the measurements.
The high trapping factor is as expected for a device on a membrane and so thermally isolated from the substrate.
A more detailed comparison is in progress.
In our view, these experimental measurements confirm that $\eta$ is indeed frequency dependent as calculated by our model.
Using the calculations of $\eta$, we may be able to differentiate between different phonon trapping factors based on the device response, or choose the phonon trapping factor to achieve the desired response.

\section{Conclusion}

We have shown that we are able to calculate the quasiparticle effective temperature resulting from uniform absorbed power using a simple analytical expression \cref{eq:analytical} in agreement with the calculated steady state nonequilibrium distributions from a full solution of the nonlinear kinetic equations, for device operating parameters (materials, temperatures, powers, and photon frequencies) typical of KIDs and similar superconducting pair breaking detectors.
The effective temperature provides a good approximation to the average quasiparticle lifetimes and surface impedance of the superconductor~\cite{Goldie2013}.
Therefore \cref{eq:analytical} allows significantly simpler inclusion of nonequilibrium effects in higher level calculations and device models.
For example, quasiparticle heating due to readout power in KIDs, which may lead to hysteresis~\cite{Thompson2013}, can be included without calculating the detailed distributions.
We also show the calculated frequency dependent response to above-gap frequency photons (particularly in the THz range), represented by the quasiparticle generation efficiency $\eta$, is in agreement with recent experimental measurements.

As a next step, we are exploring using \cref{eq:analytical} and our calculations of $\Sigma_s$ to implement a complete electrothermal model of a KID, as described in~\cite{Thomas2014pre}. This will allow exploration of device response, electrothermal feedback and hysteresis phenomena in detail.
We also note that though presented in this work in the context of KIDs, \cref{eq:analytical} is applicable to all illuminated superconducting thin films, and so is relevant to other devices, for example superconducting qubits~\cite{Dicarlo2010,Hofheinz2008,Schoelkopf2008}, resonator multiplexers for Transition Edge Sensors (TESs)~\cite{Irwin2004}, and thin film parametric amplifiers~\cite{HoEom2012}.

\printbibliography

\end{document}